# Programmable multistability for 3D printed reinforced multifunctional composites with reversible shape change

*Shanthini Puthanveetil[#], Wing Chung Liu[#], Katherine S. Riley, Andres F. Arrieta, Hortense Le Ferrand\**

S. Puthanveetil, W.C. Liu, H. Le Ferrand

School of Mechanical and Aerospace Engineering, 50 Nanyang Avenue, Nanyang Technological University, 639798 Singapore

H. Le Ferrand

School of Materials Science and Engineering, 50 Nanyang Avenue, Nanyang Technological University, 639798 Singapore

K.S. Riley, A.F. Arrieta

School of Mechanical Engineering, Purdue University, 585 Purdue Mall, West Lafayette, IN 47907, United States

**E-mail:** hortense@ntu.edu.sg

[#]shared first authorship

**Abstract**: 4D printing empowers 3D printed structures made of hydrogels, liquid crystals or shape memory polymers, with reversible morphing capabilities in response to an external stimulus. To apply reversible shape-change to stiff lightweight materials such as microfiber reinforced polymers, we developed a composite ink that can be printed using direct-ink-writing (DIW), and that exhibits multistability around its glass transition temperature. After curing at room temperature, the flat print thermally morphs into a predefined shape upon heating at an actuation temperature and cooling down. The sample can then reversibly snap between multiple stable shapes when heated above its glass transition temperature thanks to prestress-induced multistability. The key that allows thermal morphing and prestress multistability is the microstructuring of the 3D printed composites by shear-induced alignment of reinforcing microfibers. This alignment leads to local anisotropy in thermomechanical properties and the build-up of prestresses. Furthermore, the ink composition can be tuned to generate shape-dependant reversible functional properties, such as electrical conductivity. Based on finite element modelling and experimental results, the method proposed here can be used for variety of compositions and designs, for applications where stiffness, reconfigurability and shape-dependent functionalities can be exploited.

**Main Text**

1. Introduction

3D printing (3DP) is operating a paradigm shift in the way we fabricate materials and structures [1]. 3DP can now produce objects and devices on-demand and with customized designs, using a layer-by-layer deposition that reduces waste, transport, and potentially cost [2]. One interesting aspect of 3DP is the automated design capabilities where composition and architecture can be tuned locally within a macroscopic object. Going beyond 3DP, 4D printing (4DP) adds the dimension of time and endows synthetic materials with shape-adaptability in response to external stimuli such as moisture, heat, light, or pH [3–5]. Although there are numerous examples of 4DP materials, they generally rely on specific chemistry-related properties, such as shape-memory for polymers and metals, or nano-scale self-assembly for liquid crystals [5]. Moreover, many 4DP materials are also made from intrinsically soft materials like hydrogels and elastomers which may limit their applications [6]. Lightweight stiff reinforced composites, although commonly used in load-bearing parts of robots, machines and devices, generally allow complex structural change in shape only when multiple parts are joint together into actuated structures [7]. Strategies are still to be developed to endow stiff reinforced composites with intrinsic shape adaptability and reversible shape change, without the need for joints and hinges.

Successful approaches to achieving on-demand reversible bending or twisting in structures made of stiff reinforced materials exploit electrical actuation [8,9] or multistability, which is the structural characteristics of being able to rapidly snap between multiple stable states [10]. Multistability can be generated in prestressed, thin bilayer shells with reinforcements oriented in opposing directions. The morphing between the stable states is triggered by an external force that can be mechanical [11] or electromagnetic [12,13] in nature. Multistable composites have been used for robotic systems [14], aircrafts morphing wings [15] and energy harvesting systems [16]. However, these multistable structures are generally made from long fiber reinforced composites fabricated using prepregs that are handled manually and that can only provide limited shape complexity. To remediate these issues, bistability has been recently achieved in epoxy matrices reinforced with microparticles to endow fast morphing in curved samples [17–19]. To do so, magnetically responsive ceramic microplatelets were added to an epoxy matrix and oriented in specific directions using an external magnetic field before curing the matrix. To fabricate bilayer shells, sequential casting was used to obtain perpendicular microplatelets orientations between the two layers. During curing and cooling, the two layers shrank in perpendicular directions leading to the building up of internal stresses. After unmoulding, those internal stresses stored between the two layers were released and led to the deformation of the bilayer into a curved and bistable shell. This approach has been used to realize diverse functionalities into the material by optimizing the mixture composition and the composite microstructure. For example, using a combination of nickel flakes and aluminum oxide microplatelets coated with iron oxide, bistable shells with electrical conductivity and magnetic properties have been made [18]. Nevertheless, the processing method by sequential casting is tedious and would still benefit from a freeform approach to attain more shape complexity and diversity. Furthermore, although the use of magnetic fields is low cost and effective, local orientations of reinforcements in plane is yet to be demonstrated to create multistable complex shapes.

Recently, reversible morphing in 3D printed specimens of high stiffness has been realized using fused deposition modeling (FDM) [20,21]. In this process, a filament made of polylactic acid (PLA), a shape-memory polymer, was extruded through a hot nozzle. During the extrusion, the polymer chains were stretched in the direction of extrusion, resulting in printed filaments with anisotropic pre-strain. When heated above the glass transition temperature ($T_g$), the polymer chains recoiled back to their original state due to shape memory, creating directional shrinkage. Printed plates with perpendicular filament directions exhibited bistability when the specimen was placed at a temperature above $T_g$. This behaviour allowed the obtention of multiple encoded shapes within one single printed object [20,21]. To the best of our knowledge, 4D printed multistability leveraging the glass transition behavior has only been demonstrated using PLA, which is a thermoplastic with shape memory characteristics. Adapting this strategy to thermosets and composites that have no shape memory properties would broaden the range of applications of 3D printed multifunctional stiff structures that exhibit programmable and reversible morphing.

In this paper, we thus developed a composite ink system to yield a stiff epoxy reinforced by structural and functional particles, namely glass microfibers (GF), carbon nanotubes (CNTs) and carbon black (CB). This composition conferred the 3D printed parts strong mechanical properties and electrical conductivity. Direct-ink-writing (DIW), a printing method based on the extrusion of a viscous ink through a nozzle, was used for 3DP [22]. During extrusion, GF align along the print direction due to shear forces, creating local anisotropic microstructure inside the part [23,24]. We optimized the ink composition to achieve the rheological properties necessary for the extrusion and alignment of the microstructural reinforcements. Then, selected designs were 3D printed and their orthotropic properties measured. Using finite element analysis (FEA) and experiments, we demonstrated that programmable and reversible morphing can be realized in 3D printed reinforced composites using multistability from induced pre-stressing by heating them to an actuation temperature. In short, the morphing of the material follows the sequence: 3D printed flat, heating at elevated temperature above the glass temperature Tg and cooling down to induce thermal morphing, and heating around Tg for prestress multistability and reversible morphing. Furthermore, we also showed shape-dependant electrical conductivity in the samples. The process and the morphing strategy presented in this study are independent of special properties of the matrix which does not present shape memory, and is highly tuneable in terms of composition, microstructure, shape, and functional properties. Supported by FEA, it thus becomes possible to predict and design lightweight, stiff and strong structures augmented with programmable morphing, reversible shape change, and functional properties for diverse applications in areas such as robotics, aerospace, and architecture.

## 2. Experimental Section

### 2.1 Materials

The epoxy resin system (MS casting 1000 resin and MA casting 1000 hardener) was supplied by Weicon South East Asia Pte Ltd, Singapore, the CNTs (MWCNTs, Graphistrength C100) by Arkema, France, the milled glass fibers (E-glass, average length and diameter of 80-120 µm and 12-13 µm, respectively) by Nippon Electricals, Japan, and the carbon black (Printex, XE 2-B) by Degussa Gmbh, Germany. All materials were used as received. From the supplier MSDS, the epoxy curing is recommended to be carried out at an ambient temperature of 20°C. The cured resin is said to attain 90% of its maximum strength after 6 hours and maximum strength by 36 hours.

### 2.2 Ink preparation and characterisation

Resin, GF, CNTs and CB were first mixed using an overhead stirrer (Caframo Lab solutions) at 350 rpm for 10 min followed by 250 rpm for 5 min before being degassed in a vacuum chamber (Binder VD 53, Fischer Scientific Pte Ltd, Singapore). Before printing, the hardener was added to the mixture at a resin to hardener ratio of 100:20. To covert the wt% of CB and CNTs into vol%, a density of 2 g.cm$^{-3}$ was assumed. The shear profiles of the inks were measured directly after their preparation using a rheometer (Bohlin Gemini II, Germany) with a cone-plate set-up.

**2.3 3D printing**

The inks were printed using a modified version of 3D Potterbot Micro8, from 3D Potter, USA. A fixture was printed in polylactic acid using a fused deposition printer (Prusa, USA) to receive small cartridges of 5 mL (Optimum syringe barrels 7012094, Nordson EFD) to which a printing nozzle of 0.45 mm diameter (Metcal 920125-DHUV) was screwed (see SI Figure S1). The gcodes of the prints were written using the Ideamaker software (Raise3D) where input parameters were adjusted for each print. Typically, for the final ink selected, the extrusion ratio, nozzle offset distance, layer height, layer width and nozzle speed were of 2.2 (equivalent to piston speed of 0.02 mms$^{-1}$), 0.5 mm, 0.5 mm, 0.5 mm and 17.5 mms$^{-1}$, respectively. The printing was carried out on a flexible thick silicon substrate for easy removal of the sample after printing. The printed samples were then left to cure at room temperature of approximately 25°C for 48 hours to ensure complete curing in the samples.

**2.4 Thermo-mechanical properties**

The stiffness and coefficient of thermal expansion of the 3D printed samples were determined using dynamical mechanical analysis (Q-800, Texas Instrument). To obtain the material properties parallel and perpendicular to the printing direction, single-layered samples with dimensions of 10-20-0.45 mm$^3$ and 3D printed filament alignments of 0° and 90° were printed for these measurements. After curing for 48 hours, the samples were mounted onto a tensile stage. For the determination of the Young's moduli at different temperatures, the samples were first heated at selected temperatures of 25°C, 90°C, 120°C, 140°C and 160°C for 15 minutes before the stress-strain curves were recorded in isothermal conditions with a ramp force of 1 N.min$^{-1}$ up to 17 N. The coefficients of thermal expansion were determined on similar samples under a heating rate of 5 °C.min$^{-1}$ from room temperature to 160 °C using an optical dilatometer (TA instruments ODP 868). All tests were repeated on independent samples at least 5 times. Thermogravimetric analysis (TA Q500) and differential scanning calorimetry (TA Q200) were performed to study the samples' thermal stability and $T_g$.

### 2.5 Electrical properties

The local electrical conductivity of the composites was measured using a 2-points probe portable device (DT-9205A, NT) with a constant probe spacing $d$ of 1 mm. At least five measurements were averaged for each point. The conductivity was calculated using $\sigma = \frac{d}{R \cdot t \cdot l}$ with $\sigma$ the conductivity (S.mm$^{-1}$), $t$ the thickness of the sample, $l$ the width of the electrode of 0.250 mm and $R$ the measured resistance in MOhms. The colour maps were plotted using Matlab. All electrical measurements were performed at room temperature. A 4-point probe method could not be used due to the curvature of the samples, making consistent measurements with a 4-probe setup impractical.

### 2.6 Microstructural characterization

The microstructure of the printed materials was investigated using scanning electron microscopy (SEM) after gold coating (JSM-7600F, JEOL, Japan). Fiber alignment was determined by calculating the roundness of the cross-section of the glass fibers from the SEM images using ImageJ (NIH, USA) and Matlab. This roundness was calculated as $O_f = \frac{d_{1,GF}}{d_{2,GF}}$, where $d_{1,GF}$ and $d_{2,GF}$ are the lengths of the principal axes of the glass fibers cross-sections. Similarly, the roundness of the filaments was determined using ImageJ and calculated using $O = \frac{d_1}{d_2}$, with $d_1$ and $d_2$ are the principal axes of the extruded filaments.

### 2.7 Morphing

Printed bilayers were placed on a flat ceramic plate and heated in an oven from room temperature to a selected activation temperature. The samples were kept at the activation temperature for approximately 10 mins to ensure that the stresses were released. In relaxed state, the samples became flat against the ceramic plate. The heating was then switched off and the samples were left to cool back to room temperature. After cooling, the samples were imaged using a standard camera and their curvatures measured using ImageJ. The heating and cooling generated a curvature in the samples and triggered differential pre-stress and bistability as explained in Section 3.4. To observe bistability at high temperature, the samples were placed in a hot water bath at a controlled temperature of 95 °C and the change of shape between the 2 stable states was manually triggered using plastic tweezers to provide an input force. At least 3 samples of each design were fabricated and characterized for repeatability.

**2.8 Finite Element Modelling**

The samples were modelled in Abaqus 2020 using linear elastic material properties and geometric nonlinear analysis. A structured mesh composed of quadrilateral S4R shell elements was used for each structure. The experimentally measured initial dimensions and material properties as measured in Section 2.4, according to the relevant temperature (See Table S1), were input into the model. The structure was initially modelled in the as-printed flat state at the maximum heating temperature. Then the appropriate temperature change was applied, causing deflection to the first stable state. To trigger snap-through to another stable state, enforced deflections were applied to relevant corner nodes. The enforced deflections were then released, and the model assumed its closest stable state. Strain energy and reaction force plots were used to verify stable states. The curvatures were measured by exporting the stable state geometry and fitting a quadratic surface to the points in Matlab. For the model with concentric curvilinear print paths, the curvilinear reinforcements were modelled using a cylindrical material coordinate system with the appropriate origin location as the material coordinate system. Overall, the FEA followed the procedure laid out in [19–21].

## 3. Results and discussion

### 3.1 Development of the composite ink

To obtain an ink that can be printed using DIW and that yields strong mechanical properties and electrical conductivity, we chose a composite system consisting of a liquid resin reinforced with structural and functional microfibers and nanoparticles. To solve the antagonistic requirements between extrudability that demands good flowability, and strong mechanical properties that needs a high concentration of reinforcement, we selected an epoxy system with a low viscosity of 1.3 Pa.s under elevated shear strain rates in the shear-thinning regime (see Supplementary Information (SI) Figure S2 for the rheological information) and tuned the concentration of reinforcing glass microfibers (GF) (**Figure 1**). We used a fixed concentration in functional reinforcements, carbon black (CB) and multiwall carbon nanotubes (CNTs), of 0.1 and 0.5 vol%, respectively. This led to a final composite exhibiting electrical conductivity, toughness, and good processability (see SI Figure S3 for details) [25]. Although the CNTs and CB may also increase the final mechanical properties, the GF of average aspect ratio 10 were used as the main structural reinforcement. During the printing, the glass microfibers are expected to align due to the shear forces that develop at the extruding nozzle, giving the printed filament anisotropic properties (**Figure 1A**).

The optimal concentration of GF to add to the composite ink was selected based on the ink's rheological properties, namely shear-thinning and shape-retention capabilities. Increasing the volume fraction in GF from 35 to 50%, all the ink compositions exhibited shear-thinning (**Figure 1B**). Furthermore, the viscosity at rest $\eta_0$, determined in the linear viscosity range of shear-strain curves, increased with the concentration in GF, while the flow index of the inks decreased (**Figure 1C**). The flow index *n* characterizes the flowability, determined by fitting the stress-strain curves in the shear-thinning range with:

$$\tau = \eta\dot{\gamma} = k\dot{\gamma}^n + \tau_0, \qquad \text{equation (1)}$$

where $\eta$ is the viscosity, $\tau$ the shear stress, $\dot{\gamma}$ the shear strain, $k$ the flow consistency index, and $\tau_0$ a constant [26]. For printable inks, the stress should be independent of the shear rate, leading to a flow index *n*~0 [26].

The shape-retention of extruded filaments was determined by measuring their roundness $O$ after curing (**Figure 1D**). The curing of the matrix was time-dependent and started after the mixing of the components. Changes in viscosity in the reservoir of the printer could be neglected due to the fast printing speed: a 15 by 15 cm$^2$ bilayer could be printed within 10 min, which is larger than the dimensions of the samples that will be discussed in this study. We found that the highest filament roundness could be obtained with the composite ink containing 45 vol% in GF. Furthermore, the shape-retention was correlated with a pronounced alignment of the glass fibers along the printing direction, with a roundness parameter of the fiber cross-section $O_f$ of 0.9, measured at the center of the filament (**Figure 1E**). Since the shear stresses that develop during extrusion decrease with the distance from the walls of the nozzle [27], alignment at the center of the filament is representative of the alignment within the entire filament. We found that concentrations in GF ≥ 50% led to the formation of aggregates, whereas concentrations ≤ 40% led to poor filament roundness and poor fiber orientations. We measured the viscoelastic properties of the ink such as the storage and loss modulus to confirm these findings (Figure S4). As a result, the composite ink containing 45% GF was selected and used in the remainder of the paper.

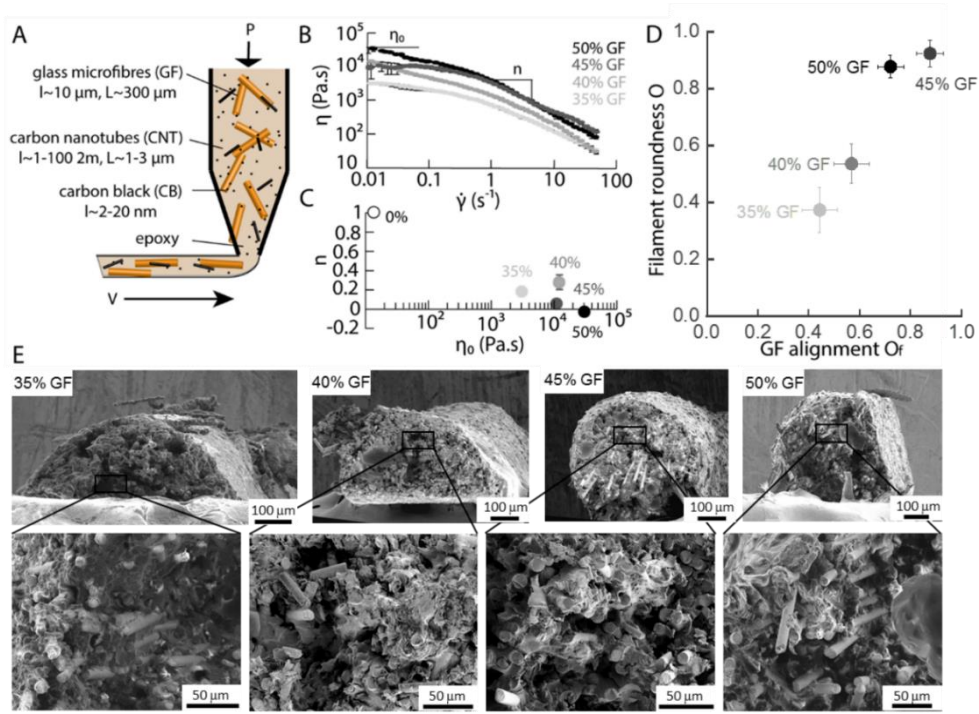

**Figure 1. Rheological properties of the ink and microstructure.** (**A**) Schematics of the ink composition and printing process. *P* and *V* represent pressure and printing velocity, respectively. (**B**) Viscosity $\eta$ as a function of the shear rate $\dot{\gamma}$ for composite inks containing 0.1% CB, 0.5% CNT, and 35 to 50% GF. $\eta_0$ is the viscosity at rest and *n* the flow index. (**C**) *n* as a function of $\eta_0$ for 35 to 50 % GF. (**D**) Roundness *O* of printed filaments as a function of the roundness $O_f$ of the cross-section of the GF, for extruded filaments from inks containing 35 to 50 % GF. (**E**) Electron micrographs of the cross-sections of extruded filaments through a nozzle of 0.45 mm diameter. Inserts are close-up views of the middle of the cross-sections.

### 3.2 3D Printing

The 3D printing of the composite ink was carried out on a extrusion printer able to apply extrusion speeds ranging between 35 to 60 mm.s$^{-1}$, typically higher than most common direct-ink-writers [22,28,29]. The printhead was modified to accommodate fine and accurate nozzle tips of 0.45 mm diameter (**Figure 2A** and SI Figure S1). The curing agent was introduced to the homogeneous mixture of resin and particles immediately before the printing. After the print, the samples were kept 2 days at room temperature to ensure total curing. The pressure drop in the conical nozzle was calculated at ~1 MPa which is in the range of pressures used in DIW (see SI for calculation details and SI Figure S5) [30–34].

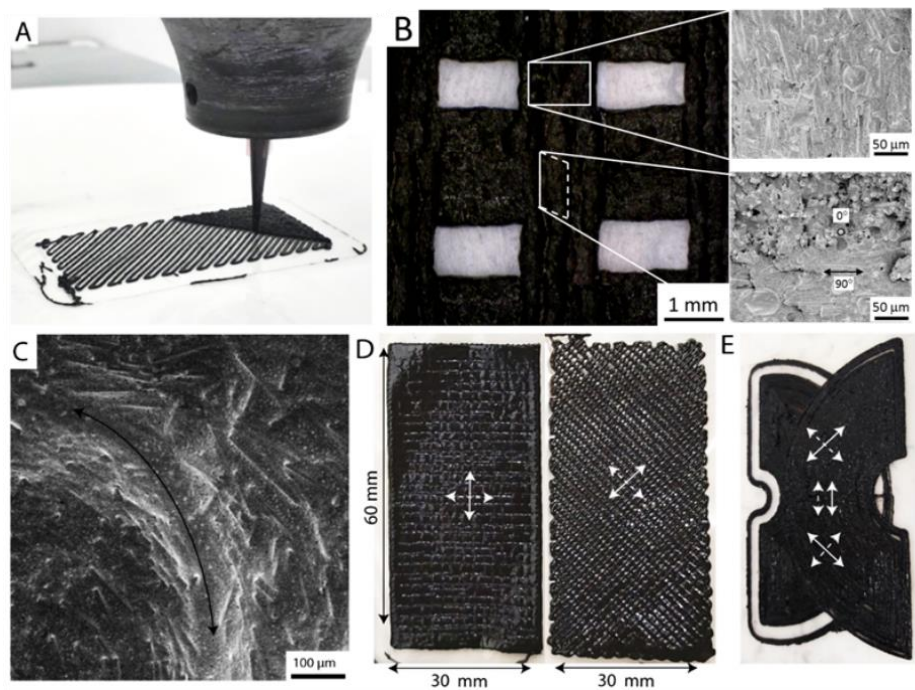

**Figure 2. Direct-ink-writing of microstructured bilayers.** (**A**) Picture of the 3DP set-up with conical nozzle of diameter 0.45 mm. (**B**) Optical image of a 3DP bilayer grid and electron micrographs of close-up views showing alignment of GF along the printing direction (insert, top) and the perpendicular orientation of the GF in the cross-section (insert, bottom). (**C**) Micrograph showing aligned glass fibers along curved directions as indicated by the arrow. (**D**) Pictures of 3DP samples with [0/90] and [-45/45] layup. (**E**) Picture of a 3DP bilayer with locally varying GF orientations along curved directions. The white arrows indicate the GF orientation directions in each layer.

Furthermore, no mechanical instabilities developed and no sharkskin was seen, despite the large amount of particles in the ink [35]. Although we previously ensured the shape retention of the inks, the printed lines showed a diameter of 0.49 ± 0.06 mm in width, which is ~9% larger than the printing nozzle, denoting a small die swell (**Figure 2B**). Rather than being a defect, this feature allowed a good bonding between the printed filaments by smoothing out their interfaces (**Figure 2B**). The glass fiber alignment along the printing direction remained well controlled in linear and curved printing directions (**Figure 2B,C**).

In view of 3DP composites with morphing capabilities, we printed bilayer structures with perpendicular directions of reinforcement [36,37]. Concretely, this meant printing bilayers with perpendicular printing directions, to yield perpendicular GF orientations (**Figure 2B, insert cross-section**). Examples of 3DP bilayers of total thickness 0.95 mm, width 30 mm and length 60 mm, and with [0/90] and [-45/45] directions of GF orientation are presented in **Figure 2D**. **Figure 2E** represents a 3D printed bilayer with curved printed lines and locally varying angle layup. Due to the orthotropic material properties of the composite, each layer will respond differently to changes in temperature, generating internal stresses and making programmed morphing possible.

### 3.3 Heat actuation and thermal of the composites

Thanks to the microstructure input in the 3DP bilayer, the printed composites exhibited controlled morphing after actuation at an elevated temperature above its glass transition temperature (**Figure 3**). The stiffness and thermal expansion coefficient of single layer specimen were measured in the directions perpendicular and parallel to the printed filaments to obtain the anisotropic properties of each individual layer. Anisotropic properties are necessary to generate the stress-induced morphing in the final bilayer structures. Indeed, the GF alignment provided orthotropic properties within each printed filament. The stiffness along the GF orientation direction was more than twice the stiffness in the perpendicular direction, whereas the opposite was measured for the coefficient of thermal expansion (**Figure 3A** and SI Figure S6 and S7). Thin plates with perpendicular directions of reinforcement are known to deform predictably upon expansion or contraction of their matrix. This bending deflection is due to the two layers' principal thermal expansion or contraction being in opposite directions [17,18,37].

We demonstrated similar deflection in our 3DP materials by heating a flat printed [0/90] layup at increasing temperatures. The morphing capability could be directly observed through the glass door of the convection oven (see SI Figure S8). Typically, the samples bent upon initial heating, then flattened as the temperature was further raised, due to the glass transition and relaxation. However, upon cooling, the sample underwent anisotropic contraction and bent again, in the opposite direction to the bending obtained during heating. The curvature increased as the temperature decreased until its stiffness was recovered and the shape "frozen". During this motion, morphing strips were able to lift up to 10 times their weight (see movie S1 for temperature-induced lifting).

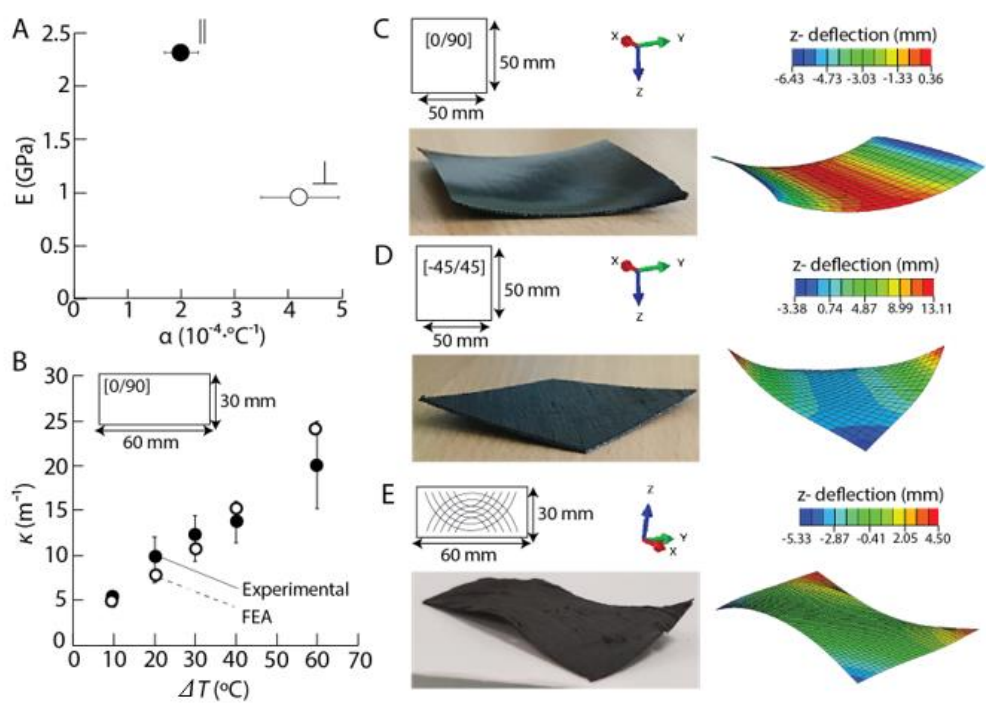

**Figure 3. Thermomechanical properties and morphing capabilities of 3D printed bilayers.** **(A)** Elastic modulus $E$ as a function of the coefficient of thermal expansion $\alpha$ in 3DP unidirectional layers parallel to the GF orientation (black) and perpendicularly (white) measured at room temperature of 25°C. **(B)** Experimental (black) and FEA (white) curvatures as a function of the degree of cooling $\Delta T = T_{max} - T_{sample}$ for [0/90] bilayer samples of dimensions 30*60*0.95 mm$^3$. The experimental curvatures are averages over 5 samples which were heated to temperatures $T_{max}$ of 120 ºC, 140 ºC and 160 ºC and various $T_{sample}$. **(C-E)** Schematics, side views and FEA models of specimens with [0/90], [-45/45] and curved layups after heating at 160 ºC and cooling to room temperature. Samples are initially flat in the xy-plane.

To take advantage of the computer-aided design capabilities of 3D printing, we conducted finite element analysis to model the deformation of our shells and compared with the experimental specimens. Predicted and experimental curvatures $\kappa$ as a function of the degree of cooling $\Delta T = T_{max} - T_{sample}$ for a rectangular [0/90] sample are presented in **Figure 3B**. Since the morphing capability is due to contractions during cooling, the resultant curvature was mainly dependent on the degree of cooling $\Delta T$, rather than the maximum heating temperature. The experimental curvatures were averaged over 5 samples heated to $T_{max}$ of 120°C, 140°C and 160°C and various $T_{sample}$ for a more accurate comparison. From **Figure 3B**, our model can accurately predict the experimental curvatures over a large range of $\Delta T$. Further investigation on the composite material's thermal stability using thermogravimetric analysis and differential scanning calorimetry also revealed that the samples could be heated up to a $T_{max}$ of approximately 200 °C without degradation or changes in the microstructure. Indeed, no change in heat flow or in weight were observed till 200 °C (Figure S9). Since the glass transition temperature is of 70 °C, the maximum $\Delta T$ possible could reach 130 °C and lead to large range of curvatures after cooling down.

**Figures 3C-E** show experimental and FEA results for [0/90], [-45/45] and curved layups after being heated at 160 ºC and cooled to room temperature (see SI Figure S10 for the morphing of the [-45/45] sample with temperature). In all cases, the FEA model was able to accurately predict the shapes. These results demonstrate that morphing of 3DP resins reinforced with microfibers can be actuated by temperature. This enables the formation of complex, 3D

structures that are initially printed flat with highly customizable, spatially distributed properties. Contrary to other morphing materials used for 3DP, the deformation does not result from intrinsic shape memory characteristics of the matrix or reorganisation of nanocrystals during phase transitions, but solely on the microstructure. Similar strategy has been applied to hydrogels but the deformation was induced by moisture and the materials exhibited weak mechanical properties [37,38]. Thus, our approach provides an exciting opportunity to create unusual 3D shapes from thin shells that also have structural properties.

**3.4 Reversible shape change through prestress-induced multistability**

3D printing of structures with encoded shape change is referred to as 4DP, as the structures are programmed to change over the fourth dimension of time in response to an external stimulus. This basic property of programmed shape change is demonstrated in Section 3.3 via thermal morphing. However, reversible shape change is more difficult to achieve. Stiff reinforced composites such as those 3D printed here, cannot reversibly change shape, in contrast to 4D printed hydrogels or smart materials [3]. We show here that reversible shape change between pre-programmed shapes can be achieved in 3DP reinforced thermosets by inducing mismatched pre-straining that in combination with the glass transition effect results in elastic bistability (**Figure 4**) [39,40]. A structure is bistable when contrasting prestresses stored in the material cause the structure to exhibit two energy minima. When an input force above a threshold magnitude is applied in the right direction, a bistable structure will undergo a rapid snap-through from one stable state to another [41,42]. Since both stable states are energy minima, no external force is necessary to maintain the structure in either state [43]. In contrast with the thermal morphing demonstrated in Section 3.3, bistability-based shape change is fast, reversible, repeatable, and elastic based on pre-stress [44–46]. Thus, pre-stress based bistability can be encoded in a structure without relying on specific material characteristics, such as shape memory, making this approach generalizable to a variety of possible material systems.

Typically, internal pre-stresses depend on the microstructure, orthotropic properties and macroscopic geometry and can be thermally induced. By cooling the structures from an elevated temperature, as shown in Section 3.3, we encode contrasting pre-stress fields in each layer. This thermal activation process at a high temperature up to 200 °C is only required to be performed once on each sample. The shape after cooling corresponds to one stable state. However, one challenge faced with reinforced epoxies is their high stiffness: although they can develop large curvatures during heating and cooling, deformation at room temperature cannot be achieved without fracture. To allow for sufficient deformability to snap between stable states, the modulus thus needs to be significantly decreased. A convenient way to realize this is by manipulating the samples around their glass transition temperature, $T_g$. The matrix used in this paper had a $T_g$ of 55 °C. After addition of the fillers, the composite's $T_g$ raised to 70 °C (see SI Figure S9). Above $T_g$, the stiffness significantly decreased from 1 GPa to 30 MPa perpendicularly to the GF and from 2.5 GPa to 80 MPa in the parallel direction (**Figure 4A**). This improved compliance allows the structures to undergo the high strains necessary for snap-through without fracture. Since the pre-stress was encoded using cooling from a much higher temperature, it is retained in the structure and enables the presence of multiple stable states. When cooled back below $T_g$, the structures return to being stiff and monostable.

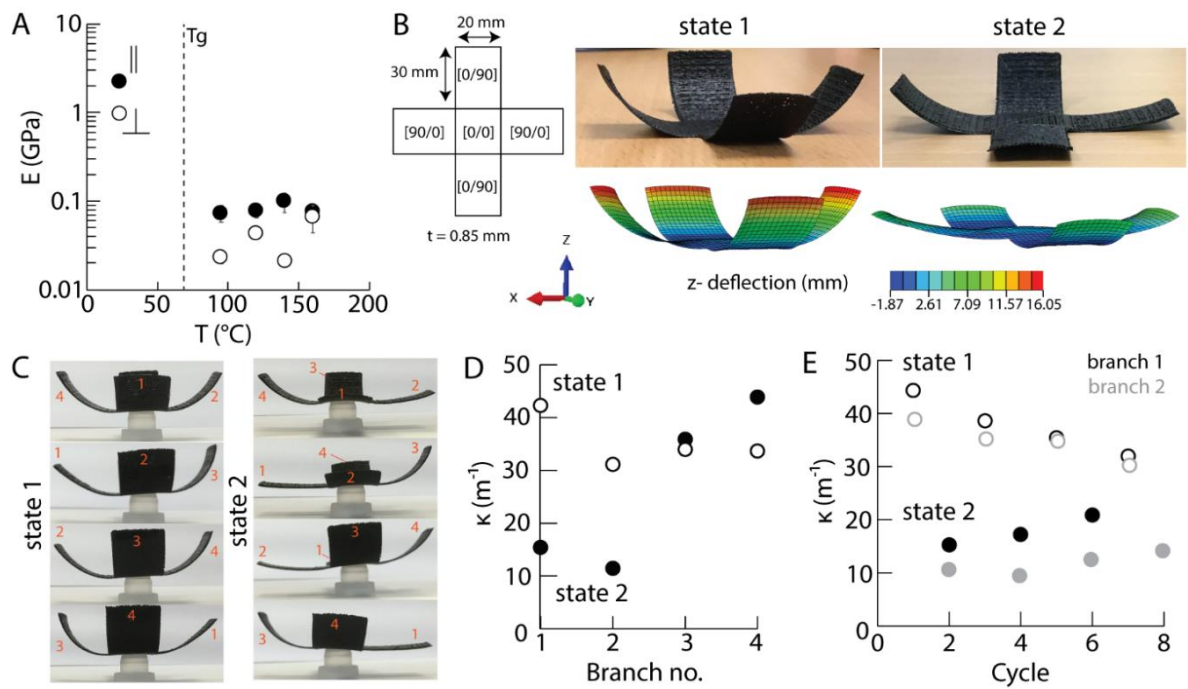

**Figure 4. High temperature multistability for reversible shape change.** (**A**) Elastic modulus *E* as a function of the temperature for 3DP samples, parallel (black) and perpendicularly to the GF alignment direction. $T_g$ is the glass transition temperature. (**B**) Schematic, pictures and FEA models of a cross with local [0-90] lay-up after activation at 160 °C. The sample could be switched reversibly between stable state 1 and stable state 2 when it was immersed into hot water above $T_g$. (**C**) Optical images of the two local stable states of individual branches of the cross. The number refers to the branch number. For each state, the sample was rotated to allow visualization of the curvatures of each branch. (**D**) Curvature of the branches in their two stable states: black, state 1 and white, state 2. (**E**) Curvature as a function of the morphing cycle between the 2 stable states at 95 °C for the branches 1 and 2 of the cross, black and grey, respectively. The empty dots represent the state 1 and the shaded dots the state 2.

To demonstrate the high temperature multistability as well as distributed reinforcement schemes realizable with 3DP, a cross with local [0/90] layups was modelled in FEA and 3DP (**Figure 4B**). The cross is a thin bilayer with local perpendicular orientations of GF, which could not be easily obtained using manufacturing methods other than 3DP. Only the middle of the cross had a symmetric [0/0] layup. After heating the flat, cured 3D printed cross to 160 °C and cooling to room temperature, the four branches morphed and curved upwards. This cooling from elevated temperature encodes contrasting thermal pre-stress fields in each layer of the cross: this is the thermal actuation step. Then, placing the curved cross in a hot water bath at 95 °C, the individual branches each exhibited two stable states (**Figure 4C**), one with higher curvature and the second with lower curvature. Since each branch is bistable, the entire cross is globally multistable as each branch can be actuated independently *via* a triggering input force (see SI Movie S2). The stable states in the branches could be preserved after removing the sample from the hot water to cool down. To simplify, we refer to state 1 the state of the cross where all the branches are in their higher curvature state, and state 2 when all the branches are in their lower curvature state. In the experimental specimen, the branches 1 and 2 displayed larger deformations between their two stable states, as compared to the branches 3 and 4. However, in the experimental model and as anticipated from the symmetry of the cross, the curvatures for branches 1 and 3 were similar, but different from the branches 2 and 4. In FEA, this difference is explained by the boundary interaction between the center of the cross with the [0/0] layup and the differently oriented unsymmetric layups of the branches. In the experiment, interfilament gaps and defects are thought to be responsible for the larger morphing of branches 1 and 2. Indeed, branches 1 and 2 had visible defects between the printed filaments, whereas branches 3 and 4 showed better filament packing (see SI Figure S11).

Furthermore, reversibility between the stable states could be repeated for at least 50 cycles in our samples while maintaining the sample at the desired temperature (**Figure 4D** and see Supplementary movie S2). However, the curvatures varied with time and the number of morphing cycles up to around the first 10 cycles, which could be due to stress relaxation. After these cycles, the sample in its state 2 was cooled to room temperature. After heating again at 95 °C, the branches in state 1 could regain their original curvatures measured at cycle 1. This indicates that no cracks or defects developed during the multiple actuation cycles. These results thus show that temperature-dependent reversible shape change between pre-programmed stable shapes can be realized in 3D printed reinforced thermosets.

**3.5 Functional properties**

In addition to being stiff and able to morph, the composites were also lightweight with a density of $1.8 \pm 0.2$ g.cm$^{-3}$ and electrically conductive. The ink composition used here was indeed designed to show electrical conductivity (see discussion in SI and Figure S3). After morphing of the 3DP composites, local strains led to local variations in electrical conductivity. It is therefore possible to design not only the 3D shapes, but also the local conductivity (**Figure 5**). Indeed, after morphing, local elements of volume are stretched to different degrees, tuning the local volume fraction of conductive elements, CB and CNTs. With an initial total volume fraction close to the percolation threshold, a small local strain would result in a measurable variation of the conductivity. With a constant ratio of CB and CNT of 0.2:1, but with a varying total volume fraction $\phi_{cond}$, the composition chosen percolated at around $\phi_p = 0.45$ vol% (**Figure 5A**). With an ink composition at $\phi_{cond}= 0.6$ vol%, a local increase in concentration due to compression would increase the conductivity, whereas a local stretching would decrease it. The conductivity varied with the concentration in CNT and CB following equation 2:

$$\sigma = C \cdot (\phi_{cond} - \phi_p)^t, \qquad \text{(equation 2)}$$

with $C$ a constant equal to $3.55 \cdot 10^{-6}$ and $t$ the percolation exponent equal to 1.232. An increase of 0.1 vol% from $\phi_{cond}$ would thus lead to an increase of $0.24 \cdot 10^{-6}$ S.mm$^{-1}$ and a decrease of 0.1 vol% to a decrease of $0.31 \cdot 10^{-6}$ S.mm$^{-1}$. The composites will show thus a higher sensitivity upon stretching than compression. To confirm this, we measured the conductivity on the top of a cured 3D printed unidirectional layer bent manually at room temperature to achieve different curvatures (**Figure 5B**). The conductivity was also found to depend upon the GF alignment with high conductivity perpendicularly to the GF. Since the CNTs were randomly oriented in the matrix (see SI Figure S12), this anisotropy must be due to the insulating properties of the GF as well as the anisotropic properties in each filament. Indeed, during curing, epoxies show a few % of shrinkage, which may have been restricted along the GF direction but allowed perpendicularly, thereby increasing the contacts between the conductive elements in that direction. Furthermore, as expected, the conductivity decreased as the bending curvature and therefore the stretching at the point of measurement increases, with a more pronounced decrease in the perpendicular direction.

The variation of the conductivity with the fibre alignment and the curvature could be exploited further for morphing-induced conductivity. This is an additional functionality that has potential interest for computational sensing where a material is able to sense, compute, and actuate, intrinsically. An example of a simple 3DP pressure sensor is presented in SI Movie S3. Although each branch of the cross exhibited a specific electrical conductivity according to its stable state and fiber orientation (**Figure 5C**), further investigations are required to match the local strain developed after morphing with the local electrical conductivity, although the decrease in electrical conductivity qualitatively matched the simulated strains (see SI text and Figure S13). It is also noted that the conductivities of our samples were generally quite low. However, this could be easily circumvented if an application requires higher conductivity, by replacing the multiwall CNTs with electrically superior single-walled CNTs, or replacing part of the milled glass fibers with conductive milled carbon fibers instead.

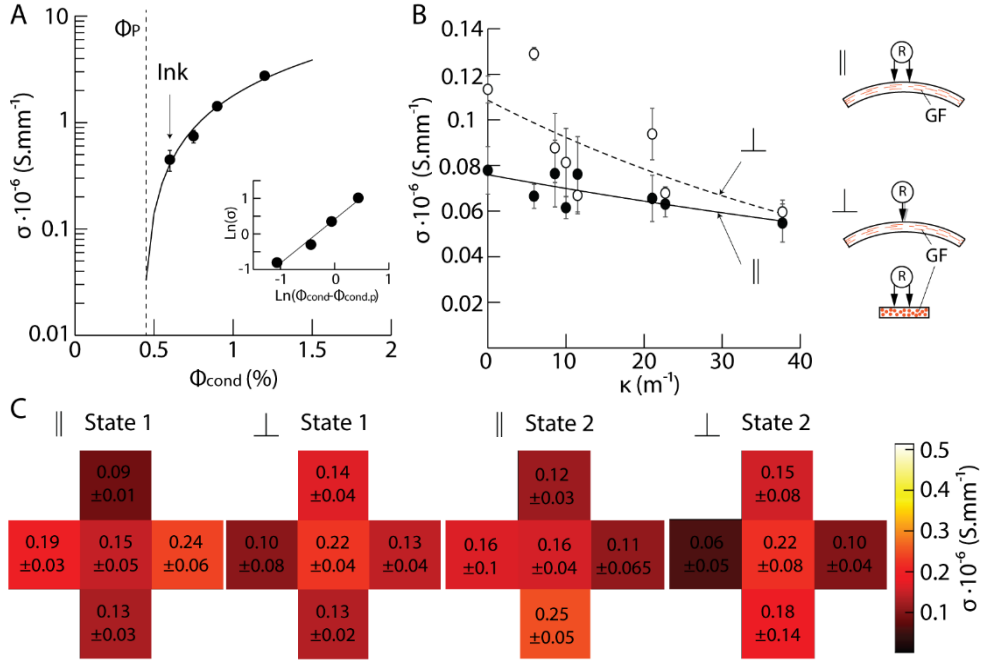

**Figure 5. Shape-dependant electrical response.** (**A**) Electrical conductivity $\sigma$ as a function of the total concentration in conductive elements $\Phi_{cond}$. The line corresponds to a fit using the percolation theory, $\Phi_p$ is the percolation threshold and ink correspond to the ink composition. (**B**) Electrical conductivity $\sigma$ as a function of the curvature of a unidirectional stripe, measured parallel and perpendicularly to the GF alignment. The schematics indicates how the resistance was measured. The lines are guide to the eyes. (**C**) Colour maps representing the local electrical conductivity at the surface of the bistable cross in its two stable states and measured parallel and perpendicularly to the GF orientation. The measurement was carries out on the inside of the cross, under compression.

## 4. Conclusions

In this paper, we have shown that stiff microfibre reinforced 3D printed composites can exhibit multistability, a type of reversible morphing, independently of any shape memory properties. Indeed, we exploited the capability of direct-ink-writing to align reinforcing microfibers through shear to build local microstructures. 3D printed thin bilayers with global or local perpendicular directions of fibre alignment could morph at temperatures above the glass transition temperature of the matrix. Using microstructural designs that encode directional, mismatched pre-stress, we establish a manufacturing process yielding bistable behaviour above $T_g$, thereby enabling reversible shape morphing between encoded states in thermoset polymers. Tuning the composition to input electrical properties, the 3D printed materials are stiff, lightweight, multistable upon heating, and with a shape-dependant electrical conductivity. The microstructuring strategy exploited in this paper recalls bio-inspired approaches to functionalize artificial materials and to combine antagonistic properties [47–50]. The principle could thus be applied to inks containing other functional elements and other microstructural designs to create structures with encoded and reversible change in shape and function for applications in aerospace to robotics.

Acknowledgements

The authors acknowledge financial support from the Ministry of Education, Singapore under Grant No. 2019-T1-001-002). This material is based upon work supported by the National Science Foundation Graduate Research Fellowship Program under Grant No. DGE-1333468. Any opinions, findings, and conclusions are those of the authors and do not necessarily reflect the views of the National Science Foundation.